\documentclass[aps,twocolumn,prd,nofootinbib]{revtex4}
\usepackage{amsmath}
\usepackage{graphicx}
\usepackage{dcolumn}
\usepackage{bm}
\usepackage{amssymb}
\usepackage{latexsym}

\begin{document}

\title{Inflation with multiple sound speeds: a model of multiple DBI type actions and non-Gaussianities}

\author{Yi-Fu Cai$^{a}$\footnote{Email: caiyf@ihep.ac.cn} and
Hai-Ying Xia$^{b,a}$\footnote{Email: haiyxia@gmail.com}}

\affiliation{${}^a$ Institute of High Energy Physics, Chinese
Academy of Sciences, P.O.Box 918-4, Beijing 100049,
P.R.China}


\affiliation{${}^b$Research Center for Eco-Environmental Sciences,
Chinese Academy of Sciences, Beijing, 100085, P. R. China}

\begin{abstract}
In this letter we study adiabatic and isocurvature perturbations
in the frame of inflation with multiple sound speeds involved. We
suggest this scenario can be realized by a number of generalized
scalar fields with arbitrary kinetic forms. These scalars have
their own sound speeds respectively, so the propagations of field
fluctuations are individual. Specifically, we study a model
constructed by two DBI type actions. We find that the critical
length scale for the freezing of perturbations corresponds to the
maximum sound horizon. Moreover, if the mass term of one field is
much lighter than that of the other, the entropy perturbation
could be quite large and so may give rise to a growth outside
sound horizon. At cubic order, we find that the non-Gaussianity of
local type is possibly large when entropy perturbations are able
to convert into curvature perturbations. We also calculate the
non-Gaussianity of equilateral type approximately.
\end{abstract}

\maketitle


\section{Introduction}

Inflationary cosmology has become the prevalent paradigm to
understand the early stage of our universe, with its advantages of
resolving the flatness, homogeneity and monopole problems
\cite{inflation_bible, Starob}, and predicts a scale-invariant
primordial power spectrum consistent with current cosmological
observations \cite{Komatsu:2008hk} very well. However, a single
field inflation model often suffers from fine tuning problems on
the parameters of its potential, such as the mass and the coupling
constant.

In recent years, people has noticed that, when a number of scalar
fields are involved, they can relax many limits on the single
scalar inflation model\cite{Liddle:1998jc}. Usually, these fields
are able to work cooperatively to give an enough long inflationary
stage, even none of them can sustain inflation separately. Models
of this type have been considered later in
Refs.\cite{Malik:1998gy, Kanti:1999vt, Copeland:1999cs,
Green:1999vv}. The main results show that both the e-folding
number ${\cal N}$ and the curvature perturbation $\zeta$ are
approximately proportional to the number of the scalars $N$.
Later, the model of N-flation was proposed by Dimopoulos {\it et
al.}\cite{Dimopoulos:2005ac}, which showed that a number of axions
predicted by string theory can give rise to a radiatively stable
inflation. This model has explored the possibility for an
attractive embedding of multi-field inflation in string theory.

Over the past several years, based on the recent developments in
string theory, there have been many studies on its applications to
the early universe in inflationary cosmology. However, people
still often encounter fine tuning and inconsistency problems when
they try to combine string theory with cosmology as reviewed in
Ref.\cite{Cline:2006hu}. Facing to these embarrassments, it is
usually suggested that N-flation is able to relax these troubles
and so can let stringy cosmology survive. A good example is that
Piao {\it et al.} have successfully applied assisted inflation
mechanism to amend the problems of tachyon
inflation\cite{Piao:2002vf}. There are also many works on
investigating multi-field inflation models in stringy cosmology,
for example see Refs. \cite{Majumdar:2003kd, Brandenberger:2003zk,
Becker:2005sg, Cline:2005ty, Gmeiner:2007uw, Ashoorioon:2009wa}.

Recently, an interesting inflation model, which has a
non-canonical kinetic term inspired by string theory, was studied
intensively in the literature. Due to a non-canonical kinetic
term, the propagation of field fluctuations in this model is
characterized by a sound speed parameter and the perturbations get
freezed not on Hubble radius, but the sound horizon instead. One
specific realization of this type of models can be described by a
Dirac-Born-Infeld-like (DBI)
action\cite{Aharony:1999ti,Myers:1999ps}. Based on brane
inflation\cite{Dvali:1998pa}, the model with a single DBI field
was investigated in detail\cite{Silverstein:2003hf, Chen:2004gc,
Chen:2005ad}, which has explored a window of inflation models
without flat potentials. In this model, a warping factor was
applied to provide a speed limit which keeps the inflaton near the
top of a potential even if the potential is steep.

In this paper, we study an inflation model involving multiple
sound speeds with each sound speed characterizing one field
fluctuation. We suggest this scenario can be realized by a number
of general scalar fields with arbitrary kinetic forms, and these
scalars have their own sound speeds respectively\cite{Cai:2008if}.
Therefore, we call this model as ``{\it Multi-Speed Inflation}".
In this model, the propagations of field fluctuations are
individual, and the usual conceptions in multi-field inflation
models might be not suitable in this scenario. For example, in a
usual generalized N-flation model, the length scale for
perturbations being freezed takes the unique sound horizon;
however, in our model it corresponds to the maximum sound horizon.

Specifically, we consider a double-field inflation model, with
each field being described by a DBI action and the total action is
constructed by the sum of they two. It is worth emphasizing that
our model is different from the usual DBI N-flation in which only
multiple moduli fields are involved in one DBI
action\cite{Ward:2007sj, Easson:2007fz, Huang:2007hh, Ward:2007gs,
Langlois:2008mn, Contaldi:2008hr, Langlois:2009ej}, but ours is
constructed by multiple DBI type actions (``DBIs"), as proposed in
Ref. \cite{Cai:2008if}. The model we considered in the paper can
be achieved as follows. We consider two D3-branes in a background
metric field with negligible covariant derivatives of field
strengths and we assume that these branes are decoupled from
others. Besides, we also need to neglect the backreaction of those
branes on the background geometry as is usually done in brane
inflation models. Specifically we are interested in two
phenomenological scenarios. The first one is these two scalars
work cooperatively like those in usual N-flation models, which
gives the predictions on primordial perturbations the similar as
those obtained in single field DBI inflation. The second one is
the scenario of cascade inflation with one scalar providing first
several efolding numbers and then the other finishing the rest.
For the second scenario, if the mass term of one field is much
lighter than that of the other, the entropy perturbation could be
quite large and so may give rise to a growth outside sound
horizon. At cubic order, we find that the non-Gaussianity of local
type is possibly large if entropy perturbations can be converted
into curvature perturbations. We also calculte the non-Gaussianity
of equilateral type approximately.

The paper is organized as follows. In Section II, we propose a
model of {\it Multi-Speed Inflation}, then study its generic
background dynamics. In Section III, we study its linear
perturbations using Arnowitt-Deser-Misner (ADM) formalism, and
show that different field fluctuations are governed by different
sound speeds respectively. In this case we provide a new
decomposition on adiabatic and isocurvature perturbations. In
Section IV, we analyze a specific inflation model constructed by
two DBI type actions, first investigate its background evolution,
and then give its curvature perturbation and entropy perturbation,
and finally its non-Gaussianity is addressed.

In the paper we take the normalization $M_p^2=1/8\pi G=1$ and the
sign of metric is adopted as $(-,+,+,+)$ in the following.

\section{The model}

Our starting point is the action as follows,
\begin{eqnarray}
S=\int d^4x\sqrt{-g} \bigg[ \frac{1}{2}R + \sum_I P_I(X_I,\phi_I)
\bigg]~,
\end{eqnarray}
with
\begin{eqnarray}
X_I\equiv-\frac{1}{2}g^{\mu\nu}\nabla_\mu\phi_I\nabla_\nu\phi_I~,
\end{eqnarray}
defined as the kinetic term of the $I$-th scalar field $\phi_I$.
This model involves multiple kessence-type fields. For simplicity,
we assume that there are no couplings between scalar fields, so
each field evolves independently except for gravity
coupling\footnote{We refer Refs. \cite{Goldwirth:1991rj,
Brandenberger:2003py} for a discussion on initial condition of
inflation.}.

An inflation model constructed with a single kessence was
originally proposed by \cite{ArmendarizPicon:1999rj} and later its
perturbation theory has been studied\cite{Garriga:1999vw}. In the
literature this type of model has been widely studied, and one of
the most significant features is that there is an effective sound
speed describing the propagation of the perturbations
\cite{ArmendarizPicon:2003ht, Piao:2007eq, Magueijo:2008pm,
Khoury:2008wj}. However, one may already notice that in our model,
for each a kessence field there is one sound speed
correspondingly. Therefore, the field fluctuations in our model do
not propagate synchronously. In the current paper, our main
interests focus on the effects of multiple sound speeds in
perturbation theory. However, before studying the perturbations,
we first take an investigation on the background equations.

By varying the action with respect to the metric, we can obtain
the energy-momentum tensor of the form
\begin{eqnarray}
T^{\mu\nu}=\sum_I\bigg(P_Ig^{\mu\nu}+{P_I}_{,X_I}\nabla^\mu\phi_I\nabla^\nu\phi_I\bigg)~,
\end{eqnarray}
where ${P_I}_{,X_I}$ denotes the partial derivative of $P_I$ with
respect to $X_I$. Moreover, the scalar fields satisfy generalized
Klein-Gordon equations, which are given by
\begin{eqnarray}
\nabla_\mu({P_I}_{,X_I}\nabla^\mu\phi_I)+{P_I}_{,I}=0~,
\end{eqnarray}
where ${P_I}_{,I}$ is the partial derivative of $P_I$ with respect
to the scalar $\phi_I$.

Considering a spatially flat Friedmann-Robertson-Walker (FRW)
spacetime with its metric
\begin{eqnarray}
ds^2=-dt^2+a^2(t)d\vec{x}^2~,
\end{eqnarray}
we can read the energy density and pressure of a field $\phi_I$
from the energy-momentum stress
\begin{eqnarray}
\rho_I=2X_I{P_I}_{,X_I}-{P_I}~,~~p_I=P_I~.
\end{eqnarray}
The equations of motion for the scalar fields reduce to,
\begin{eqnarray}\label{kg}
\ddot\phi_I+(3H+\frac{\dot{P_I}_{,X_I}}{{P_I}_{,X_I}})\dot\phi_I-\frac{{P_I}_{,I}}{{P_I}_{,X_I}}=0~,
\end{eqnarray}
where we define the Hubble parameter $H\equiv\dot a/a$. Moreover,
we introduce the sound speed parameters,
\begin{eqnarray}
{c_s^2}_I\equiv\frac{{p_I}_{,X_I}}{{\rho_I}_{,X_I}}=\frac{{P_I}_{,X_I}}{{P_I}_{,X_I}+2X_I{P_I}_{,X_IX_I}}~,
\end{eqnarray}
and also use the dimensionless parameters
\begin{eqnarray}
s_I\equiv\frac{\dot{c_s}_I}{H{c_s}_I}~,
\end{eqnarray}
for each of which measures the variation of the sound speed
${c_s}_I$ in one Hubble time.

\section{Linear perturbations}

Now we start to study the linear perturbations of the model
introduced in the previous section. Since we are working in the
frame of a cosmological system, the metric perturbations ought to
be included as well as the field fluctuations. However, one can
eliminate one degree of freedom by taking a suitable gauge. We
would like to expand the action to the second order with the ADM
formalism\cite{Arnowitt:1962hi}. In this formalism, we can
eliminate one extra degree of freedom of perturbations at the
beginning of the calculation, by choosing the spatially flat gauge
with the spatial curvature vanishing as $^{(3)}R=0$.

\subsection{ADM formalism and the action at quadratic order}

To start, the spacetime metric in the ADM formalism is written as,
\begin{eqnarray}
ds^2=-N^2dt^2+h_{ij}(dx^i+N^idt)(dx^j+N^jdt)~,
\end{eqnarray}
with $N$ being the lapse function and $N_i$ the shift vector.
Substituting this metric into the original expression of the
action, we get
\begin{eqnarray}
S=\int dt dx^3 \sqrt{h} \bigg[ N\sum_IP_I +
\frac{1}{2N}(E_{ij}E^{ij}-E^2) \bigg]~,
\end{eqnarray}
where $h=det(h_{ij})$, and the tensor $E_{ij}$ is defined as
\begin{eqnarray}
E_{ij}=\frac{1}{2}(\dot h_{ij}-\nabla_iN_j-\nabla_jN_i)~,
\end{eqnarray}
which is related to the extrinsic curvature of the spatial slice
with $K_{ij}=N^{-1}E_{ij}$. We use the lowercase index $i$ to
denote the spatial coordinates.

To vary the action with $N$, we obtain the Hamiltonian constraint,
\begin{eqnarray}\label{constraint1}
2\sum_IP_I-\frac{1}{N^2}(E_{ij}E^{ij}-E^2+2\sum_I{P_I}_{,X_I}v_Iv_I)=0~,
\end{eqnarray}
where there is
\begin{eqnarray}
v_I=\dot\phi_I-N^i\partial_i\phi_I~;
\end{eqnarray}
while the variation of the action with respect to $N_i$ yields the
momentum constraint as follows,
\begin{eqnarray}\label{constraint2}
\nabla_j\bigg[ N^{-1}(E_i^j-E\delta_i^j) \bigg] =
N^{-1}\sum_I{P_I}_{,X_I}v_I\partial_i\phi_I~.
\end{eqnarray}

We have already restricted ourselves to the spatially flat gauge
of the FRW background with the spatial metric as
$h_{ij}=a^2(t)\delta_{ij}$. Therefore, the degrees of freedom
merely comes from the field fluctuations as the following
decomposition,
\begin{eqnarray}
\phi_I(t,\vec{x})\rightarrow\phi_I(t)+\delta\phi_I(t,\vec{x})~.
\end{eqnarray}
Meanwhile, we need to expand the lapse function and shift vector
in form of,
\begin{eqnarray}
N=1+\alpha~,~~N_i=V_i+\partial_i\beta~,
\end{eqnarray}
where the scalar functions $\alpha$ and $\beta$ can be expressed
in terms of the field fluctuations $\delta\phi_I$, and $V_i$
belongs to the vector modes and so can be eliminated to second
order.

By solving the linearized constraint equations (\ref{constraint1})
and (\ref{constraint2}), we have
\begin{eqnarray}
\alpha=\frac{1}{2H}\sum_I{P_I}_{,X_I}v_I\delta\phi_I~,
\end{eqnarray}
and
\begin{eqnarray}
\partial^2\beta &=& \frac{a^2}{2H} \sum_I \bigg[
-\frac{{P_I}_{,X_I}}{{c_s}_I^2}v_I\delta{v_I} +
({P_I}_{,I}-2X_I{P_I}_{,IX_I})\delta\phi_I
\nonumber\\
&&+\frac{{P_I}_{,X_I}}{H}(\frac{X_I{P_I}_{,X_I}}{{c_s}_I^2}-3H^2)v_I\delta\phi_I
 \bigg]~,
\end{eqnarray}
and here $\delta{v_I}=\delta\dot\phi_I$ to linear order.

Making use of the above results, now we can expand the action to
quadratic order. To do some integrations by parts and regroup the
terms, the second order action takes the form
\begin{eqnarray}\label{S2}
S_2 &=& \int dtdx^3 \frac{a^3}{2} \sum_I \bigg[
\frac{{P_I}_{,X_I}}{{c_s}_I^2}\delta\dot\phi_I^2
-\frac{{P_I}_{,X_I}}{a^2}\partial_i\delta\phi_I\partial_i\delta\phi_I \nonumber\\
&&+2{P_I}_{,IX_I}\dot\phi_I\delta\phi_I\delta\dot\phi_I
-\sum_JM_{IJ}\delta\phi_I\delta\phi_J \bigg]~,
\end{eqnarray}
where the effective mass matrix of the field fluctuations is given
by
\begin{eqnarray}
M_{IJ} &=&
\frac{1}{2H}(\dot\phi_I^2{P_I}_{,IX_I}\dot\phi_J{P_J}_{,X_J}
+\dot\phi_I{P_I}_{,X_I}\dot\phi_J^2{P_J}_{,JX_J})\nonumber\\
&&+\frac{1}{4H^2}\sum_K(1-\frac{1}{{c_s}_K^2})\dot\phi_K^2{P_K}_{,KX_K}{P_I}_{,X_I}{P_J}_{,X_J}\nonumber\\
&&-\frac{1}{a^3}\frac{d}{dt}
[\frac{a^3}{4H}(2+\frac{1}{{c_s}_I^2}+\frac{1}{{c_s}_J^2})\dot\phi_I{P_I}_{,X_I}\dot\phi_J{P_J}_{,X_J}]\nonumber\\
&&-{P_I}_{,II}\delta_{IJ}~,
\end{eqnarray}
which is strongly suppressed by slow-roll parameters in the frame
of usual inflationary cosmology.

One may notice, if there is only one field, the above results are
consistent with a model of single kessence field as analyzed in
Ref \cite{Garriga:1999vw}. However, in our case there are multiple
sound speeds which govern the propagations of the field
fluctuations. For each sound speed, there is a critical length
scale which takes the form $c_s/H$. We would like to call this
scale as {\it sound horizon}.

\subsection{Curvature perturbations and isocurvature perturbations}

Having obtained the second order action, we can proceed to study
the field fluctuations by solving their perturbation equations in
concrete models. However, before doing that, let us take a closer
look at the kinetic terms of the field fluctuations. From Equation
(\ref{S2}), it is not straightforward how to define the modes of
curvature and isocurvature perturbations, since the model involves
more than one sound speeds which make the usual decomposition of
an orthonormal basis in field spaces invalid. So we need to
develop a new decomposition which should includes the information
of the sound speeds. To do it, we need to go back to the basic
definition of curvature perturbation in perturbed Einstein's
equations.

A widely used quantity characterizing the gauge invariant
curvature perturbation is given by
\begin{eqnarray}
{\cal R} &\equiv&
\Phi-H\frac{\delta{q}}{\rho+p}\nonumber\\
&=&\Phi_k+H\frac{\delta\sigma_k}{\dot\sigma}~.
\end{eqnarray}
Here $\Phi$ being the gravitational potential, $\delta{q}$
represents the perturbation of momentum, and $\sigma$ is the
so-called adiabatic field.

One can find that this field takes the form
$\dot\sigma=\sqrt{\sum_I\dot\phi_I^2}$ in usual cases. However,
recall that, in our model the $(0i)$ components of the perturbed
energy-momentum stress give the momentum perturbation as follows,
\begin{eqnarray}
\delta{q} = - \sum_I {P_I}_{,X_I} \dot\phi_I \delta\phi_I~;
\end{eqnarray}
while, the background energy density and the pressure yield
\begin{eqnarray}
\rho+p = \sum_I {P_I}_{,X_I} \dot\phi_I^2~.
\end{eqnarray}
Thus the adiabatic field in our model is given by
\begin{eqnarray}
\dot\sigma=\sqrt{\sum_I{P_I}_{,X_I}\dot\phi_I^2}~,
\end{eqnarray}
and its perturbation can be expressed as
\begin{eqnarray}
\delta\sigma = \sum_I
{P_I}_{,X_I}\frac{\dot\phi_I}{\dot\sigma}\delta\phi_I~,
\end{eqnarray}
which characterize the adiabatic fluctuations.

Moreover, we usually define another useful gauge-invariant
variable, curvature perturbation on uniform-density
hypersurface\cite{Bardeen:1983qw}, with its expression as follows,
\begin{eqnarray}
\zeta \equiv -\Phi-H\frac{\delta\rho}{\dot\rho}~.
\end{eqnarray}
On large scales, we have ${\cal R}\simeq-\zeta$ in a spatially
flat universe. Thus, both two quantities can be used to describe
adiabatic fluctuations. If the matter content of a cosmological
system is made of multiple components, we can define the curvature
perturbation associated with each individual energy density
components, which are given by
\begin{eqnarray}
\zeta_I = -\Phi-H\frac{\delta\rho_I}{\dot\rho_I}~.
\end{eqnarray}
Since in a system with multiple matter components, there are
non-vanishing entropic pressure perturbations even with every
component being adiabatic. So we can describe the entropy
perturbations by using the following expressions,
\begin{eqnarray}
{\cal S}_{IJ}=3(\zeta_I-\zeta_J)~,
\end{eqnarray}
which are so called relative entropy perturbations.

Furthermore, we can define the adiabatic unit vector
$e^I_{\sigma}$ as follows,
\begin{eqnarray}
e^I_{\sigma}=\sqrt{{P_I}_{,X_I}}\frac{\dot\phi_I}{\dot\sigma}~.
\end{eqnarray}
To take a further step, we have
\begin{eqnarray}
\delta\sigma=\sqrt{{P_I}_{,X_I}}\delta\phi_Ie^I_{\sigma}~,
\end{eqnarray}
and thus we can see that this decomposition is no longer on an
orthonormal basis.

In order to make the analysis more explicitly, we consider an
example of two kessence fields. In this case the adiabatic
perturbation and entropy perturbation can be given by
\begin{eqnarray}
\delta\sigma&=&\cos\theta\sqrt{{P_1}_{,X_1}}\delta\phi_1+\sin\theta\sqrt{{P_2}_{,X_2}}\delta\phi_2~,\\
\delta{s}&=&-\sin\theta\sqrt{{P_1}_{,X_1}}\delta\phi_1+\cos\theta\sqrt{{P_2}_{,X_2}}\delta\phi_2~,
\end{eqnarray}
and the angle is defined by
\begin{eqnarray}
\tan\theta=\frac{\sqrt{{P_2}_{,X_2}}\dot\phi_2}{\sqrt{{P_1}_{,X_1}}\dot\phi_1}~.
\end{eqnarray}
Using the above decomposition, we now obtain the formal
expressions of dimensionless curvature and isocurvature
perturbation variables,
\begin{eqnarray}\label{pertRS}
{\cal R}\simeq H\frac{\delta\sigma}{\dot\sigma}~,~~{\cal
S}=H\frac{\delta{s}}{\dot\sigma}~,
\end{eqnarray}
on the spatially flat slices. Note that, as ${\cal S}$ is not
directly observable during inflation, what we are interested in is
its spectral index but not the amplitude. Therefore, its
normalization is quite arbitrary. We take such a particular choice
in Eq. (\ref{pertRS}) since it has been widely used in usual
double-field inflation as shown in Refs.
\cite{Langlois:1999dw,Gordon:2000hv,Bartolo:2001rt,Wands:2002bn}.

\section{A model of two DBI type actions}

In this section, we study a specific inflation model involving
multiple sound speeds. The model was originally proposed in Ref.
\cite{Cai:2008if}, where the model is constructed by multiple DBI
type actions and its general feature has been studied under
certain approximations. Now we focus on a concrete model which
only involves two DBI type actions, with
\begin{eqnarray}
P_I(X_I,\phi_I)=\frac{1}{f(\phi_I)}[1-\sqrt{1-2f(\phi_I)X_I}]-V_I(\phi_I)~,
\end{eqnarray}
with $I=1,2$. This model might be viewed as an effective
description of D-brane dynamics (for example see Refs.
\cite{Maldacena:1997re}). Considering a system constructed by two
D3-branes in a background metric field with negligible covariant
derivatives of the field strengths and assuming that these two
branes are falling into their own throats, this system can be
described by the above action which has a stringy origin as shown
in Ref. \cite{Taylor:1999pr}.

In this model, the scalar $\phi_I$ describes the position of the
brane. If we consider the branes are falling into the AdS-like
throats and neglect the backreaction of the branes upon the
background geometry, the warping factor usually takes the form
\begin{eqnarray}
f(\phi_I)=\frac{\lambda_I}{\phi_I^4}~.
\end{eqnarray}
This assumption can be satisfied when the contribution of the
background flux is much larger than that from the branes.

\subsection{Background equations}

After having introduced the model, now we can study its background
dynamics in the frame of FRW metric. For the scalars, the sound
speeds are given by
\begin{eqnarray}
{c_{s}}_I=\sqrt{1-2f(\phi_I)X_I}~,
\end{eqnarray}
and there is ${P_I}_{,X_I}=1/{c_s}_I$. The above equation yields
\begin{eqnarray}
|\dot\phi_I|=\phi_I^2(\frac{1-{c_s}_I^2}{\lambda_I})^{\frac{1}{2}}~.
\end{eqnarray}
Since in the current paper we focus our interests on the
relativistic limit with small sound speeds, then
$|\dot\phi_I|\simeq\phi_I^2/\sqrt{\lambda_I}$.

Specifically, we consider the case of IR type potential with the
form of
\begin{eqnarray}
V_I= V_{0I}-\frac{1}{2} m_I^2 \phi_I^2 ~.
\end{eqnarray}
The first part of the potential $V_{0I}$ origins from the
anti-brane tension from other throats. In IR DBI
inflation\cite{Chen:2005ad}, D-branes roll towards the tip of the
throats, thus the potential contains terms like tachyon. Moreover,
due to the warping factor $f(\phi_I)$, those scalars are able to
stay near the top of their potentials, and so we have $H^2\simeq
\frac{1}{3}\sum_I V_{0I}$. In the following, we would like to
investigate the background in details.

In order to obtain a semi-analytic solution, we would like to take
a useful assumption with $s_I\equiv\frac{\dot c_{sI}}{Hc_{sI}}$
being small numbers, and consider the relativistic limit of the
branes.

A similar case of a single field has been studied in
\cite{Chen:2005ad}. However, in our model the total lagrangian is
constructed by two DBI fields, where each one field contribute one
lagrangian and has its own sound speed respectively. Therefore, we
actually have two sound speeds. Recall the generalized
Klein-Gordon equations (\ref{kg}), they can be reexpressed as
follows,
\begin{eqnarray}\label{phieq2}
\frac{d}{dt}(\frac{\dot\phi_I}{c_{sI}}) +
3H\frac{\dot\phi_I}{c_{sI}} +
\frac{f_{,I}}{f^2}(1-c_{sI})-\frac{f_{,I}\dot\phi_I^2}{2fc_{sI}}+V_{,I}=0~.
\end{eqnarray}
Under the relativistic limit of the scalars we can have an ansatz,
which takes the following form,
\begin{eqnarray}
\phi_I=-\frac{\sqrt{\lambda_I}}{t} (
1-\frac{\alpha_I}{(-t)^{p_I}}+... )~,
\end{eqnarray}
where we set $t\rightarrow-\infty$ at the beginning of
inflation\footnote{In stringy configuration, the flux-antibrane
annihilation in the multiple throats naturally provides an
attractive point for small field inflation with the branes
generating at the tips of the throats through tunneling from an
eternal inflation\cite{Chen:2004gc,Chen:2006hs}.}. Therefore, to
insert the ansatz into the above equation, then we find the
leading terms in Eq.(\ref{phieq2}) come from the second term
\begin{eqnarray}
\frac{3H\sqrt{\lambda_I}}{\sqrt{2\alpha_I(p_I-1)}(-t)^{2-\frac{p_I}{2}}}
\end{eqnarray}
and the potential term which is equal to
\begin{eqnarray}
\frac{\sqrt{\lambda_I}m_I^2}{t}~.
\end{eqnarray}
The others are suppressed by $\frac{1}{Ht}$ which is negligible in
inflation (where $|Ht|\gg1$ or equivalently
$\phi_I\ll\sqrt{\lambda_I}H$), and this requirement is consistent
with the assumption that the scalars lie on the top of potential
during inflation. Finally, by matching the leading terms, we get
$p_I=2$ and $\alpha_I=\frac{9H^2}{2m_I^4}$, and so the solutions
of the scalars are given by
\begin{eqnarray}
\phi_I\simeq-\frac{\sqrt{\lambda_I}}{t}(1-\frac{9H^2}{2m_I^4t^2}+...)~.
\end{eqnarray}
Making use of the solutions, we have the sound speeds
\begin{eqnarray}\label{soundspeed}
{c_s}_I\simeq\frac{3H}{m_I^2t}~,
\end{eqnarray}
to the leading order, and so can check that the approximations
with $s_I\ll1$ are consistent with the equations of motion when
$t\rightarrow -\infty$.

\subsection{Quantum fluctuations and power spectrum}

Now let us study the dynamics of quantum fluctuations in this
model. To do so, we go back to the second order action (\ref{S2})
directly and define the new variables which are canonically
normalized with conformal time. These variables are defined as
\begin{eqnarray}
v_I=a\frac{\sqrt{{P_I}_{,X_I}}}{{c_s}_I}\delta\phi_I~,
\end{eqnarray}
for the two scalars. Thus the dominant terms of second order
action can be given as follows,
\begin{eqnarray}
S_2 \supseteq \int d\tau dx^3 \sum_{I=1}^2 \bigg[
{v_I}'^2-{c_s}_I^2\partial_iv_I\partial_iv_I+\frac{z_I''}{z_I}{v_I}^2
\bigg]~,
\end{eqnarray}
under an assumption of weakly coupling between two fields. In the
above action, we have introduced some background dependent
functions
\begin{eqnarray}
z_I=a\frac{\dot\phi_I\sqrt{{P_I}_{,X_I}}}{{c_s}_IH}~.
\end{eqnarray}

In the inflationary background, the equations of motion describing
these two canonical perturbation variables in Fourier space are
given by
\begin{eqnarray}
{v_I}_k''+({c_s}_I^2k^2-\frac{z_I''}{z_I}){v_I}_k=0~,
~~\frac{z_I''}{z_I}\simeq\frac{2}{\tau^2}~.
\end{eqnarray}
One can see that, for each field fluctuation, there is a
corresponding sound horizon respectively. This scenario is quite
different from the case considered in Ref \cite{Langlois:2008mn}
where there is only sound speed characterize the propagation of
the adiabatic mode.

To complete the quantization of the field fluctuations, we can
decompose the variables as
\begin{eqnarray}
v_I(\tau,\vec{x}) = \int\frac{dk^3}{(2\pi)^{\frac{3}{2}}} \bigg[
{a_I}_{\vec{k}}{v_I}_{k} + {a_I}_{-\vec{k}}^{\dag}{v_I}_{-k}^*
\bigg] e^{i\vec{k}\vec{x}}~,
\end{eqnarray}
where the operators $a_I$ and $a_I^\dag$ are annihilation and
creation operators, which satisfy the following commutation
relation
\begin{eqnarray}\label{creatanihi}
[{a_I}_{\vec{k}},{a_I}_{\vec{k}'}^\dag]=\delta(\vec{k}-\vec{k}')~.
\end{eqnarray}
Moreover, the normalization conditions require
\begin{eqnarray}
{v_I}_k{{v_I}_k^*}'-{v_I}_k^*{v_I}_k'=i~.
\end{eqnarray}

Eventually, we choose that all the modes of perturbations behave
as in the adiabatic Minkowski vacuum initially, and thus obtain
the solutions
\begin{eqnarray}
{v_I}_k=\frac{e^{-i{c_s}_Ik\tau}}{\sqrt{2{c_s}_Ik}}(1-\frac{i}{{c_s}_Ik\tau})~.
\end{eqnarray}
These solutions imply that the power spectrum of the field
fluctuations are of the value,
\begin{eqnarray}\label{deltaphi}
\delta\phi_I=\sqrt{P_{\delta\phi_I}}\simeq\frac{H_{*I}}{2\pi}~,
\end{eqnarray}
after they escape out of their own sound
horizons\cite{Liddle:2000cg}. The subscript `$*I$' denotes the
sound horizon crossing time for the field perturbation
$\delta\phi_I$.

Since in inflation the Hubble parameter is almost unchanged, and
the amplitudes of field perturbations are also nearly conserved
due to the feature of nearly scale-invariance, we can conclude
that the amplitudes of all field fluctuations take almost the same
value outside of their sound horizons. Due to this quantity, we
can take the maximum of those sound horizons as the final freezing
scale for all the field fluctuations, which corresponds to the
perturbation mode with the largest sound speed. For example, if we
take $m_1<m_2$ in our model which gives ${c_s}_1>{c_s}_2$
according to Eq. (\ref{soundspeed}), then the critical sound
horizon takes $\frac{{c_s}_1}{H}$.

Based on the above analysis and making use of Eqs. (\ref{pertRS}),
(\ref{deltaphi}) and the background solutions, one obtains the
curvature perturbation at the {\it maximum sound horizon} crossing
time
\begin{eqnarray}
{\cal R} \simeq \frac{{\cal N}^2}{2\pi\sqrt{\lambda_I}}
(1+\frac{27H^4}{2m_1^2m_2^2{\cal N}^2})~,
\end{eqnarray}
where we have considered the next-to-leading correction and
introduced the efolding number ${\cal N}\equiv\int Hdt$. The
entropy perturbation can be resolved as well,
\begin{eqnarray}
{\cal S} \simeq
\frac{27H^4(m_1^2-m_2^2)}{4\pi\sqrt{\lambda_I}m_1^3m_2^3}~,
\end{eqnarray}
which is proportional to a parameter, {\it relative sound speed}
$\Delta c_s$, at the crossing time, with its definition in form
of,
\begin{eqnarray}\label{rcs}
\Delta{c_s}={c_s}_1-{c_s}_2\simeq\frac{3H(m_2^2-m_1^2)}{m_1^2m_2^2t}~.
\end{eqnarray}

We can see that, if there is only one single field, the entropy
perturbation vanishes. Even for the case of double fields, the
entropy perturbation is still not large in usual DBI inflation,
since its amplitude is suppressed by the relative sound speed for
which we have expanded the detailed formalism in our concrete
model. However, one may notice that a relatively small mass term
may uplift the entropy perturbations. For example, if we take
$m_1\ll m_2$, the amplitude of entropy perturbation takes an
approximate form as ${\cal S}\sim\frac{H^4}{m_1^3m_2}$ which can
be very large due to an enough light mass $m_1$. This feature
might be very important and could be applied in a curvaton model
with DBI type actions which will be discussed in near
future\cite{Pi:2009a}.

\begin{figure}[htbp]
\includegraphics[scale=0.8]{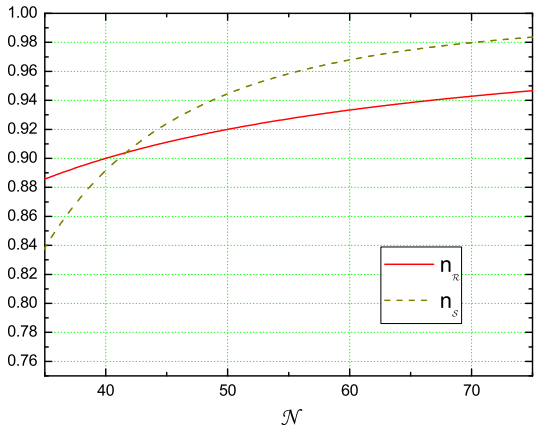}
\caption{The plot of spectral indices as functions of the
e-folding number ${\cal N}$ for curvature and entropy
perturbations. The red solid line denotes the spectral index of
curvature perturbation; the yellow dashed line denotes the
spectral index of entropy perturbation. The background parameters
are taken as:
$\lambda_1=\lambda_2=10^{14},~m_1=10^{-6},~m_2=5\times10^{-6}$.}
\label{fig:ns}
\end{figure}

Moreover, the leading term of the curvature perturbation is
consistent with the result obtained in Ref. \cite{Chen:2005ad},
and the second order contribution is suppressed by a square of the
efolding number and so we can neglect it in usual case. One should
keep in mind that, if we tune one of the masses to be small
enough, the second order term could also dominate over.
Furthermore, we give the spectral tilts as follows,
\begin{eqnarray}
&n_{\cal R}-1 \equiv \frac{d\ln P_{\cal R}}{d\ln k} \simeq
-\frac{4}{{\cal N}}~,&\\
&n_{\cal S}-1 \equiv \frac{d\ln P_{\cal S}}{d\ln k} \simeq
-8\epsilon~,&
\end{eqnarray}
in which we have the following relation
\begin{eqnarray}
\epsilon = -\frac{\dot H}{H^2}
\simeq\frac{\lambda_I(m_1^2+m_2^2)}{3{\cal N}^3M_p^2}~,
\end{eqnarray}
in our model. The dependence of the spectral indices on the
e-folding number ${\cal N}$ for different background values are
plotted in Fig. \ref{fig:ns}.

From the above figure we can read that, both the spectral index of
curvature perturbation and that of entropy perturbation are
roughly scale-invariant, but with their tilts a little red.
Moreover, the deviation of the entropy perturbation from a
scale-invariant spectrum is smaller than that of the curvature
perturbation at the regime of large efolding numbers, while larger
at small efolding numbers. According to the recent cosmological
observation, for instance WMAP five year
data\cite{Komatsu:2008hk}, we have a constraint on the amplitude
of the curvature perturbation ${\cal R}\lesssim4.8\times10^{-5}$.
So if we choose the background parameters $\lambda_I=10^{14}$, the
best fit value for efolding number is ${\cal N}\simeq 55$ which
satisfies the current cosmological observations very well.

\section{Non-Gaussianities}

In the above section we have calculated the primordial
fluctuations in linear order. If our model indeed makes sense to
the physics of the early universe, it would be necessary to extend
the theoretical framework beyond the leading order. In particular,
the information of non-Gaussianity provides a potentially powerful
discriminant between numerous models describing early universe and
have attracted considerable interests. Non-Gaussianities in usual
single field inflation models were considered in
\cite{Allen:1987vq, Salopek:1990jq, Falk:1992sf, Komatsu:2001rj,
Acquaviva:2002ud} and in more detail in \cite{Maldacena:2002vr}
and it was found that non-Gaussianities would be small. Later,
large non-Gaussianity of specific shape\cite{Babich:2004gb} which
is of equilateral type can be obtained in single DBI
inflation\cite{Alishahiha:2004eh}, but its value of local type is
still very small\cite{Chen:2006nt}. The non-Gaussianity of local
type can be sizable in bounce cosmology as studied by Ref.
\cite{Cai:2009fn}. We refer Ref. \cite{Bartolo:2004if} for a
comprehensive review on this issue and Refs. \cite{NGrecent} for
some recent developments.

In this section we use $\delta{\cal N}$ formalism
\cite{Sasaki:1995aw, Starobinsky:1986fxa, Lyth:2004gb} to study
the non-Gaussianities in details. To start, we define the power
spectrum $P_{\cal{R}}$ and bispectrum $B_{\cal{R}}$ as follows,
\begin{eqnarray}
\langle {\cal R}_{k_1} {\cal R}_{k_2} \rangle &=& (2\pi)^3
\delta^3(\vec{k}_1+\vec{k}_2) P_{\cal{R}}(k_1)~,\\
\langle {\cal R}_{k_1} {\cal R}_{k_2} {\cal R}_{k_3} \rangle &=&
(2\pi)^3 \delta^3(\vec{k}_1+\vec{k}_2+\vec{k}_3) \nonumber\\
&&\times B_{\cal{R}}(k_1,k_2,k_3)~,
\end{eqnarray}
and then these two spectra can be related in terms of the
nonlinearity parameter $f_{NL}$,
\begin{eqnarray}\label{fnlbasic}
B_{\cal{R}}(k_1,k_2,k_3) = \frac{3}{10}(2\pi)^4\frac{\sum
k_i^3}{\prod k_i^3}P_{\cal R}^2 f_{NL}(k_1,k_2,k_3)~,
\end{eqnarray}
in momentum space.

The concept of $\delta{\cal N}$ formalism identifies the curvature
perturbation with the perturbation of local expansion ${\cal
R}=\delta{\cal N}$, and so the curvature perturbation can be
expanded as follows,
\begin{eqnarray}\label{deltaN}
{\cal R} = \sum_I{\cal
N}_{,I}\delta\phi_I+\frac{1}{2}\sum_{IJ}{\cal
N}_{,IJ}\delta\phi_I\delta\phi_J+\cdots.
\end{eqnarray}
If we have calculated the three point correlators of field
fluctuations, the non-linearity parameter can obtained by making
use of above equations.

\subsection{Equilateral type}

Specifically, now we make a rough study on the three point
correlator and non-Gaussiantiy of equilateral type. A simple way
to investigate the three point correlator is to perturb the second
order lagrangian as shown in Eq. (\ref{S2}). With an assumption of
weak coupling between the fields, we perturb the sound speeds in
the quadratic lagrangian and then obtain the lagrangian with the
leading order terms up to cubic parts,
\begin{eqnarray}
{\cal L}_3 \supseteq \sum_I\frac{a^3}{2{c_s}_I^5\dot\phi_I}
[\delta\dot\phi_I^3-\frac{{c_s}_I^2}{a^2}\delta\dot\phi_I(\nabla\delta\phi_I)^2]~.
\end{eqnarray}
Correspondingly, the dominant terms in the interaction Hamiltonian
in Fourier space are given by
\begin{eqnarray}\label{Hint}
H_{int} \supseteq \int dk^3 \bigg[
-\sum_I\frac{a^3}{2{c_s}_I^5\dot\phi_I}
(\delta\dot\phi_I^3+\frac{{c_s}_I^2}{a^2}k^2\delta\dot\phi_I\delta\phi_I^2)
\bigg].
\end{eqnarray}
Then we decompose the field fluctuations in canonical quantization
process with creation and annihilation operators defined in Eq.
(\ref{creatanihi}),
\begin{eqnarray}\label{fieldpert}
\delta{\phi_I}_{k}(t)=u_I(\vec{k}){a_I}_k+u_I^*(-\vec{k}){a_I}_{-k}^\dag~,\nonumber\\
u_I(\vec{k})=\frac{H}{\sqrt{2k^3}}(1+i{c_s}_Ik\tau)e^{-i{c_s}_Ik\tau}~.
\end{eqnarray}
Since we have obtained the interaction Hamiltonian and the modes
of the field fluctuations, now we are able to calculate the three
point correlator, which takes\footnote{For simplicity, we have
assumed there is no coupling between two fields and calculate the
three-point function of each field. The studies on N-flation with
single DBI action were intensively studied in Refs.
\cite{Huang:2007hh, Langlois:2008wt, Arroja:2008yy,
Langlois:2008qf, RenauxPetel:2008gi}.},
\begin{eqnarray}
\langle
\delta{\phi_I}_{k_1}\delta{\phi_I}_{k_2}\delta{\phi_I}_{k_3}
\rangle = -i \int dt \langle
[\delta{\phi_I}_{k_1}\delta{\phi_I}_{k_2}\delta{\phi_I}_{k_3},H_{int}]
\rangle~.\nonumber\\
\end{eqnarray}
From Eqs. (\ref{Hint}) and (\ref{fieldpert}), we can see that
$H_{int}\sim \sum_I 1/{c_s}_I^{2}$. Note that this is consistent
with the result in usual single DBI inflation which is
proportional to $1/c_s^2$\cite{Chen:2006nt}.

Furthermore, from Eq. (\ref{NI}), we can read the partial
derivative of the efolding number with respect to each field.
Substituting this equation into Eq. (\ref{deltaN}) and using the
expression (\ref{fnlbasic}), we finally obtain an approximate form
of $f_{NL}$ of equilateral type as follows,
\begin{eqnarray}
f_{NL}^{equil} \sim
\frac{({c_s}_1+{c_s}_2)({c_s}_1^5+{c_s}_2^5)}{{c_s}_1^2{c_s}_2^2({c_s}_1^2+{c_s}_2^2)^2}~.
\end{eqnarray}
This is also consistent with the case of single DBI
inflation\cite{Chen:2006nt}. Moreover, if we take ${c_s}_1$ larger
than ${c_s}_2$, then the non-linearity parameter takes
$f_{NL}\sim1/{c_s}_2^2$. Therefore, we can conclude the
non-Gaussianity of equilateral type is sensitive to the smallest
sound speed (To make a comparison, the linear curvature
perturbation strongly depend on the largest sound speed, which
corresponds to the maximum sound horizon).

\subsection{Local type}

To take a further step, we consider non-Gaussianity of local type
in the model we have studied in previous section.

As analyzed previously, for all field fluctuations, their
amplitudes take almost the same value
$\delta\phi_I\simeq\frac{H}{2\pi}$ after they exit their own sound
horizons. Thus the non-linearity parameter of local type is given
by
\begin{eqnarray}\label{fnldN}
f_{NL} \simeq \frac{5}{6}\frac{\sum_{JK}{\cal N}_{,J}{\cal
N}_{,K}{\cal N}_{,JK}}{(\sum_I{\cal N}_{,I}{\cal N}_{,I})^2}~.
\end{eqnarray}
Moreover, from the formula of curvature perturbation
$(\ref{pertRS})$, we can read the following relations,
\begin{eqnarray}\label{NI}
{\cal N}_{,1} \simeq
\frac{\sqrt{\lambda_I}H\phi_1^2}{\phi_1^4+\frac{{c_s}_1}{{c_s}_2}\phi_2^4}~,~{\cal
N}_{,2} \simeq
\frac{\sqrt{\lambda_I}H\phi_2^2}{\phi_2^4+\frac{{c_s}_2}{{c_s}_1}\phi_1^4}~.
\end{eqnarray}
Moreover, we define a parameter
\begin{eqnarray}\label{qpara}
q\equiv\frac{\phi_2}{\phi_1}~,
\end{eqnarray}
which represents the ratio between two scalars. Using this
parameter we can simply describe two background evolutions. One is
$q\sim1$, which denotes that both two fields are rolling down
along their potentials synchronously as considered in previous
sections; the other is $q\gg1$, which describes that, the heavy
field $\phi_2$ enters its warping throat and provides first few
efolds for inflation and then the light field $\phi_1$ starts the
second episode of inflation. The latter case provides a specific
realization of cascade inflation\cite{Silk:1986vc, Adams:1991ma,
Polarski:1992dq, Feng:2003zua, Ashoorioon:2006wc, Bean:2008na}.

Substituting the above relations into Eq. (\ref{fnldN}) and using
the definition (\ref{qpara}), we can obtain an approximate form of
non-Gaussianity of local type directly,
\begin{eqnarray}
f_{NL}^{local}\sim\frac{q}{{\cal N}_1}~,
\end{eqnarray}
where ${\cal N}_1$ denotes the efolds contributed by the field
$\phi_1$. From this result, one can see that the non-Gaussiantiy
of local type is suppressed by the efolding number. For example,
we consider two fields are rolling down in the same rate and thus
there is ${\cal N}\simeq60$, we obtain $f_{NL}^{local}\sim
O(10^{-2})$. In this case the deviation of curvature perturbation
from Gaussian distribution is very small, and also consistent with
the result in single DBI inflation. However, if we let the light
field only provides the last several efolds during which there is
$q\gg1$, it gives $f_{NL}^{local} \propto q$ which could be very
large. We understand its physics in the following. Initially, the
heavy field $\phi_2$ dominates the background evolution, and so
the perturbations from the other field $\phi_1$ contribute on
iso-curvature modes. Later the first period of inflation ceases
with a cutoff $\phi_2\sim M_p$ and then $\phi_1$ dominates over,
and this process converts entropy perturbations into adiabatic. In
this process there is usually a non-linear growth of curvature
perturbations outside sound horizon. This mechanism has been
widely applied in curvaton \cite{Mollerach:1989hu, Linde:1996gt,
Lyth:2001nq, Lyth:2005fi, Huang:2008ze, Li:2008fma}, and Ekpyrotic
models \cite{Khoury:2001wf, Koyama:2007if, Buchbinder:2007at,
Lehners:2007wc}.



\section{Conclusion and discussions}

In this paper, we have studied an inflation model involving
multiple sound speeds. This scenario can be embedded into
inflation models with multiple components, for example a model of
N-flation, which in usual can avoid some difficulties of single
field inflation models, and so is regarded as an attractive
implementation of inflation. In recent years, there have been many
works around the issue of N-flation, such as Refs.
\cite{Kim:2006ys, Piao:2006nm, Olsson:2007he, Choi:2007fya,
Panotopoulos:2007pg, Battefeld:2008py, Ashoorioon:2008qr}, and we
refer to Refs. \cite{Wands:2007bd, Langlois:2008sg} for recent
reviews. However, in all these pioneer works, there is only one
sound speed involved. In the model we proposed, due to a number of
sound speeds, it may lead to plentiful interesting phenomena. In
the current paper, we only focus on the curvature and entropy
perturbations and has already found that the decomposition on
these two modes in usual N-flation models cannot be applied in our
model.

Specifically, we have studied a new model in which a collection of
two DBI fields drives inflation simultaneously. This model was
originally proposed in Ref. \cite{Cai:2008if}, which has
discovered that some non-perturbative effects are involved when we
study background evolution and curvature perturbation. In the
current paper, we considered a tachyonic potential and provided
detailed calculations on perturbations. To linear order, we find
that the perturbations do not get freezed until the modes exit the
maximum of all the sound horizons, and so the linear perturbations
depend on the largest sound speed. Moreover, both the curvature
perturbations and entropy perturbations are nearly scale-invariant
with red tilts. However, their spectral indices are different. For
curvature perturbations, the tilt is suppressed by the efolding
number as $1/{\cal N}$; while the tilt of the spectral index of
entropy perturbations is suppressed by efolding number as $1/{\cal
N}^3$.

Furthermore, we have investigated the non-Gaussianities of
equilateral type and local type in this model. By calculating the
three point correlators, our results show that, for the
non-Gaussianity of equilateral type is much sensitive to the
smallest sound speed which takes the form $f_{NL}\sim1/c_s^2$.
Besides, since our model is constructed by double fields, there is
entropy perturbations generated during inflation. Therefore, when
the entropy perturbations contribute to curvature perturbations at
late times of inflation, they can lead to a sizable
non-Gaussiantiy of local type. However, if both two field evolve
in the same rate during inflation, the non-Gaussianity of local
type is still suppressed by slow roll parameter.

\textbf{Acknowledgments} We are graceful to Niayesh Afshordi,
Bruce Bassett, Robert Brandenberger, Bin Chen, Hasson Firouzjahi,
Xian Gao, Ghazal Geshnizjani, Bin Hu, Miao Li, Shun-Pei Miao, Shi
Pi, Yun-Song Piao, Richard Woodard, Wei Xue, and Xinmin Zhang for
useful discussions. We specially thank Xingang Chen and Yi Wang
for valuable comments on the manuscript. We wishes to thank the
KITPC for hospitality during the program ¡°Connecting Fundamental
Physics with Cosmological Observations¡±. This work is supported
in part by National Natural Science Foundation of China under
Grant Nos. 10533010 and 10675136 and by the Chinese Academy of
Science under Grant No. KJCX3-SYW-N2.

\end{document}